\documentclass[pre,aps,showpacs,showkeys,groupedaddress]{revtex4}
\usepackage{graphicx}
\usepackage{amsmath}
\usepackage[bookmarksnumbered=true,linkbordercolor={0.8 0.8 
0.8},citebordercolor={0.6 0.7 0.6}]{hyperref}

\begin{document}

\title{Bhattacharyya statistical divergence  of quantum observables}
\author{V. Majern\'{\i}k}
\email{fyziemar(at)savba.sk}
\affiliation{Institute of Mathematics, Slovak Academy of
Sciences,  Bratislava, \v Stef\'anikova  47,
 Slovak Republic}
\author{S.Shpyrko}
\email{serge\_shp(at)yahoo.com}
\affiliation{
Department of Optics, Faculty of Science, Palack\'y University,
T\v r. 17. listopadu 50, CZ-77207 Olomouc, Czech Republic} 
\affiliation{
 Institute of Nuclear Research, Ukrainian Academy of
Sciences, pr. Nauki 47 Kiev, Ukraine }

\begin{abstract}

In this article we exploit the Bhattacharyya statistical divergence
to determine the similarity  of probability distributions of quantum
observables. After brief review of useful characteristics of the
Bhattacharyya divergence we apply it to determine the similarity of
probability distributions of two non-commuting observables. An
explicit expression for the Bhattacharyya statistical divergence is
found for the case of two observables which are the x- and
z-components of the angular momentum of a spin-1/2 system. Finally,
a note is given of application of the considered statistical
divergence to the specific physical measurement.

\end{abstract}


\maketitle

\section{Introduction}
One of the important problems in the probability theory is to find
an appropriate measure of the difference or the statistical {\it
divergence} of two probability distributions $P$ and $P'$. This
measure quantifies the degree of the similarity between them. In the
mathematical statistics the divergence of two probability
distributions is introduced as follows: If $[X,P] $ and $[X,P']$ are
two probability spaces then the so-called (Csisz\'ar's) f-divergence
of probability distributions $P$ and $P'$ is given as
\[ D_f(P;P')=\sum\limits_{x\in X} P'(x)f\left[\frac{P(x)}{P'(x)}\right]\,,\]
where $f(u)$ represents a convex function in the interval $(0,\infty)$ and
strictly convex for $u=1$ \cite{VA}.

Among the existing divergence measures of two discrete
probabilities, $P\equiv [p_1,p_2,\dots p_n]$ and $Q\equiv
[q_1,q_2,\dots q_n]$, the Kullback-Leibler statistical divergence
\cite{KU}
\[D_K(P:Q)=\sum_{i=1}^n p_i\log \left(\frac{p_i}{q_i}\right)\,,\]
is perhaps best known and most widely used. This is why this measure
has several desirable properties, such as nonnegativity and
additivity, which are crucial in its applications. $D(P;Q)$ is {\it
not}  symmetrical regarding the exchange of $P$ and $Q$. For
$D_K(P;Q)$, the inequality holds
\[\sum_{i=1}^n p_i\log{\frac{p_i}{q_i}}\geq 0\,.\]
The minimum of $D_K(P;Q)$ is obtained iff $p_i=q_i$ (see, e.g.
\cite{AD}).

Apart of the Kullback-Leibler statistical divergence, a number of
other
 divergence measures, depending on certain parameters, have been proposed and intensively
 studied by
R\'enyi \cite{RE}, Kapur \cite{KA}, Kullback and Leiber \cite{KL},
Havrda and Charvat \cite{HCH}, Tsallis \cite{TS}, \cite{T}. Some of
them satisfy the convexity condition only for restricted values of
the corresponding parameters.

However, the Kullback-Leibler, R\'enyi, Havrda-Charvat, Tsallis and
the trigonometrical \cite{MA} statistical divergences require
generally $p_i=0$ whenever $q_i=0$. From point of view of their
application, this is not a desirable property because just such
situations we often encounter in the theoretical physics especially
in statistical and quantum physics.

In the next sections, we exploit one of the first statistical
divergence measure, that was proposed in the literature, the
Bhattachryya divergence of $P$ and $Q$
which is symmetrical regarding the exchange of $P$ and $Q$ and does
not suffer from the above-mentioned shortcoming.
We attempt to apply the Bhattacharyya statistical divergence to the
quantification of the degree of similarity of two quantum
observables.

\section{The Bhattacharyya statistical divergence}

The Bhattacharyya statistical divergence of the discrete probability
distributions
\[P=p_1,p_2,\dots,p_n;\quad
Q=q_1,q_2,\dots,q_2\] is defined as \cite{BH}
\[S(P,Q)=
\sum_{i=1}^{n} (p_{i}q_{i})^{1/2}.\quad \eqno(1)\]
This divergence has the following properties:\\
(i) It becomes its maximal value equal to $1$ when the probability
distributions $P$ and $Q$ are identical.\\
(ii) Its minimal value is zero when the components of $P$ and $Q$
do not overlap.\\
(iii) Its value lies in the interval $ [0,1]$ and expresses the
degree how much the probability distributions of $P$ and $Q$ are
similar.\\
(iv) It is {\it symmetrical} regarding the exchange of $P$ and
$Q$.\\
(v) $S(P,Q)$ satisfies the properties of nonnegativity,
finiteness and boundedness.\\
(vi) It can be straightforward extended for more  than two
probability distributions \cite{MAJ}.

The Bhattacharyya statistical divergence of $P$ and $Q$ has a simple
geometrical interpretation. Consider the following vectors $\hat
P_1\equiv \{\sqrt p_1, \sqrt p_2,\dots,\sqrt p_n\}$ and $\hat Q\equiv
\{\sqrt q_1,\sqrt q_2,\dots \sqrt q_n\}$. According to Eq. (1), the
similarity measure of $\hat P$ and $\hat Q$ is simply the scalar
product of $\hat P$ and $\hat Q$ in $\bf R^{(m)}_{+}$. Since $\hat
P$ and $\hat Q$ represent the unit vectors in ${\bf R_+^{(m)}}$ its
scalar product is equal to the cosine of angle between $\hat P$ and
$\hat Q$ which, of course, has the properties (i)-(iv).

\section{Application of the Bhattacharyya statistical divergence to quantum mechanics}

Consider two observables $A$, $B$ with Hermitian operators $\hat A$,
$\hat B$ in an N-dimensional Hilbert space, whose corresponding
complete orthonormal sets of eigenvectors $\{|x^{(A)}_i\rangle\},$
$\{|x_i^{(B)}\rangle\},\dots$ $(i=1,2,\dots,N)$ are disjointed and have
nondegenerate spectra. Let $|\phi\rangle$ be a normalized state
vector of N-dimensional Hilbert space then it holds
\[|\phi\rangle =\sum_i^N a_i|x_i^{(A)}\rangle, \qquad |\phi\rangle =
\sum_j^N b_j |x^{(B)}_j\rangle,\dots\]

 Accordingly, the components of the probability distributions $P(A)$ and $P(B)$
 associated with the observables $A$ and $B$
 are \[P_i(A)=|a_i|^2=|\langle x_i^{(A)}| \phi \rangle |^2;
 \quad
 \sum_{i=1}^n P_i(A)=\sum_{i=1}^n|a_i|^2=1\quad\eqno(2a)\]
 \[ P_i(B)
=|b_i|^2=|\langle y_j^{(B)}|\phi \rangle|^2;\quad
\sum_{i=1}^nP_i(B)=\sum_{i=1}|b_i|^2=1\,.\quad \eqno(2b)\] Inserting
Eqs.(2a) and (2b) into Eq.(1) we get
\[S(P(A),P(B))=
\sum_{i=1}^n({P_i(A)P_i(B)})^{1/2}=\sum_{i=1}^n|a_i||b_i|\,.\] 

If
$A=B$ then $|a_i|=|b_i|$ for $i=1,2,...,n$ and Eqs.(2a) and (2b)
becomes
\[S(P(A),P(B))=\sum_{i=1}^nP_i(a)=\sum_{i=1}^n|a_i|^2=\sum_{i=1}^n
P_i(B)=\sum_{i=}^n |b_i|^2=1\,.\] 
Hence, for $ A\equiv B$ it follows
$S(P(A),P(B))=1$.  Given the state vector $|\phi \rangle$ and
operators $\hat A,\hat B$ the considered statistical divergence of
their probability distributions can be generally determined. To each
operator, a ray in the Hilbert space can be assigned. The quantity
$S(P(A),P(B))$ gives the closeness of different rays in Hilbert
space. If these rays are identical then their Bhattacharyya
divergence $S(P(A),P(B))$ is equal to $1$. If they are perpendicular
to each other then $S(P(A),P(B))$ becomes zero. Generally,
$S(P(A),P(B))$ $A\equiv\hat B$ and the corresponding rays of these
operators in Hilbert space are identical, i.e. the cosine of angle
between them is equal to $1$. Therefore, $S(P(A),P(B))=1$.

Next, we consider  the case of two {\it non-commuting} observables.
Consider two observables $A$ and $B$ with non-commuting Hermitian
operators $\hat A$ and $\hat B$ in an N-dimensional Hilbert space,
whose corresponding complete orthonormal sets of eigenvectors
$\{|x_i\rangle\},$ $\{y_i\rangle\}$ $(i=1,2,...,N)$ are disjointed
and have nondegenerate spectra. Let $|\phi\rangle$ be a normalized
state vector of N-dimensional Hilbert space then it holds
\[|\phi\rangle =\sum_i^N a_i|x_i\rangle, \qquad |\phi\rangle =
\sum_j^N b_j |y_j\rangle\,.\]
According the quantum transformation theory we have
\[|\phi\rangle = \left(\sum_i^N a_i \langle x_i|y_1 \rangle\right)\, |y_1\rangle +
 \left(\sum_i^N a_i \langle x_i|y_2 \rangle\right)\, |y_2\rangle +\dots=
 \sum_j^N \sum_i^N a_i \langle x_i|y_j \rangle |y_j\rangle
\]
\[P_i(A)=|\langle x_i| \phi \rangle |^2=|a_i|^2,\quad Q_j(B)
=|b_j|^2=|\langle y_j|\phi \rangle|^2=
 \Big|\left(\sum_i a_i \langle y_i|x_j \rangle\right) \Big|^2\,,\quad\eqno(3)\]
where $\langle x_i|y_j\rangle\quad i,j=1,2,3,...,N$ are the elements of
the transformation matrix ${\bf T}$ between the observables $A$ and $B$
\[{\bf T}=\left(\begin{matrix}  \langle x_1|y_1\rangle &
\langle x_1|y_2\rangle & \dots & \langle x_1|y_n\rangle \\
& \dots \\ \langle x_n|y_1\rangle & \dots & & \langle x_n|y_n\rangle
\end{matrix}\right)\]

 Inserting Eq.(3) into Eq.(1) we get for the Bhattacharyya
 statistical divergence of the probability distributions $P(A)$ and
 $Q(B)$ the expression
\begin{multline}
S(P(A),Q(B))=
\sum_{i=1}^n\sqrt{P_i(A),Q_i(B)}=
\sum_{i=1}^n|a_i||b_i|=
\sum_{i=1}^n|\langle x_i|\phi
\rangle||\langle y_i|\phi\rangle|= 
\sum_{j=1}^{n}|a_j|\,\Big|\left(\sum_{i=1}^{n}a_i\langle y_i|x_j\rangle
\right)\Big|. 
\end{multline}

Given the state vector $|\phi \rangle$ and the components of
${\bf T}$, the divergence of the probability distributions of $A$
and $B$ can be generally determined.

 Next, we present an example for
determining of Bhattacharyya divergence of two concrete {\it
complementary} observables describing a simple quantum system.

\section{An example}

 For
the sake of simplicity, we will consider the complementary
observables in a two-dimensional Hilbert space. Such system
represents a particle with spin $\hbar /2$ \cite{ME}. Determining
the probability distributions of the components $J_x$ and $J_z$ we
can calculate the similarity measure of their probability
distributions.

The state vector of this quantum system is
spinor
\[|\Psi \rangle = \left( \begin{array}{c}
                a_{1}\\
                a_{2}\\
                \end{array} \right)\,,\]
where
\[a_{1}a_{1}^{*}+a_{2}a_{2}^{*} = 1\,.\]

Its wave functions in z-representation takes the form $|\Psi
\rangle_{z} = a_{1}|z_{1}\rangle +a_{2}|z_{2}\rangle$. According
Eq.(3), the probability  $P_{z_1}$ and $P_{z_2}$ that $J_z$ is
projected on $|z_1\rangle$  and $|z_2\rangle$ is
$|a_1|^2=a_1a_1^{*}$
 and $|a_2|^2=a_2a_2^{*}$, respectively, so the corresponding
probabilistic schema becomes
\begin{center}
\begin{tabular}{c||c|c}
$J_{z}$ & $|z_{1}\rangle$ & $|z_{2}\rangle$ \\
\hline
$P_{J_z}$ & $a_{1}a_{1}^{*}$ & $a_2a_{2}^{*}$ \\
\end{tabular}
\end{center}
Using the transfer transformation matrix
\[{\bf T}=\frac{1}{\sqrt 2}\left(\begin{matrix}  1,\quad  1 \\
 1, -1 \end{matrix} \right)\]
we obtain  $ \Psi$ in its x-representation $|\Psi \rangle_{x} =
{\displaystyle\frac{a_{1} + a_{2}}{\sqrt{2}}}\,|x_{1}\rangle +
{\displaystyle\frac{a_{1}-a_{2}}{\sqrt{2}}}\,|x_{2}\rangle\,.$ Similarly, the
probabilistic scheme for $J_x$ turns out to be
\begin{center}
\begin{tabular}{c||c|c}
$J_{x}$ & $|x_{1}\rangle$ & $|x_{2}\rangle$ \\
\hline $P_{J_X}$ & $2^{-1}(a_{1}+a_{2})(a_{1}^{*}+a_{2}^{*})$ &
$2^{-1}(a_1 - a_{2})(a_{1}^{*} -
a_{2}^{*})$ \\
\end{tabular}.
\end{center}
Now, we express $a_1$ and $a_2$ by means of new
variables $r$ and $\varphi$ in the following way
\[a_{1} = \sqrt{r}\exp (i\varphi_1 ), \qquad a_{2} =
(\sqrt{1-r})\exp (i\varphi_2 ), \qquad \varphi = \varphi_{1} -
\varphi_{2}\,.\] In these variables, we obtain for $J_x$ and $J_z$ the
following probability distributions
\[ Q(J_x)\equiv \{r^2,(1-r)^2\}\]
and
\[ P(J_z)\equiv \left\{\frac{1}{2}(1+2\sqrt{r-r^2}\cos \varphi,
\frac{1}{2}(1-2\sqrt{r-r^2}\cos \varphi\right\}\,.\] The similarity measure
of these probability distributions consists of two terms
\[S(Q(J_x),P(J_z))=T_1+T_2,\quad\eqno(4)\]
where \centerline{$T_1=\sqrt{\frac{r}{2}(1+2\sqrt{r-r^2} \cos
\varphi)}$  and
    $T_2 =\sqrt{\frac{(1-r)}{2}(1-2\sqrt{r-r^2}\cos \varphi)}.$}
If $J_x$ occurs in one of its eigenstates, i.e. $r=1$ or $0$, then
the first or second term in Eq.(4) becomes zero and we obtain the
minimal value of $S(P(J_x),Q(J_y))$ equal to $T_1=T_2=\sqrt{1/2}$.
$S(P(J_x),Q(J_z))$ attains its maximal value equal to $1$ for
$r=\sqrt{1/2}$ and $\varphi=\pi/2.$ The 3D-plot $S(P(J_x),Q(J_z))$
is given in Fig. 1.

\begin{figure}[ht]
\includegraphics[scale=0.75,clip=true]{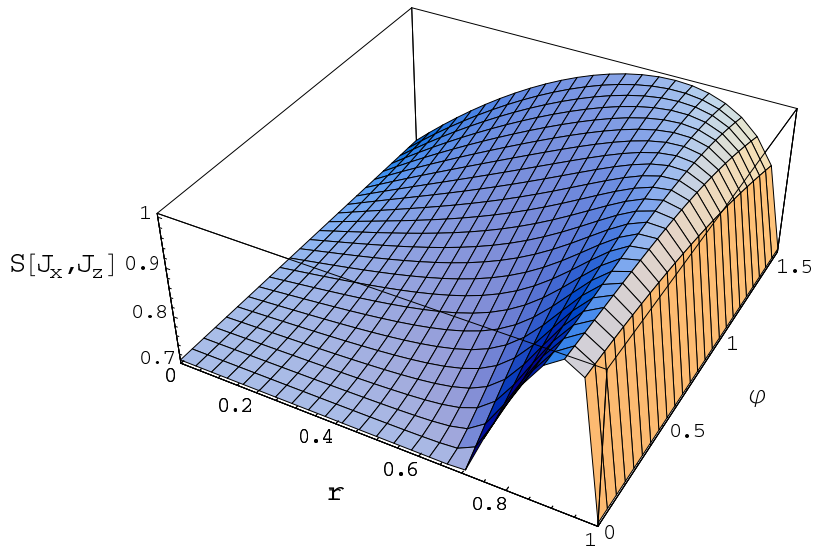}
\caption{3D-plot of the similarity measure of the probabilities
${\bf P_{J_x}}$ and ${\bf P_{J_z}}$ as function of $r$ and
$\varphi$.} \label{fig1}
\end{figure}

This graph shows that in the vicinity of $r=0.85$ and for $\varphi
\in [0, \pi/2]$ a hump occurs, where the probability distributions
are almost similar. $S(P(J_x),Q(J_z))$ never drops under the value
$\sqrt{1/2}$, therefore it holds
\[S(P(J_x),Q(J_z))\geq \sqrt{1/2}\,.\]
While S(P:Q) for two commuting observables in two-dimensional
Hilbert space can attain arbitrary value, the similarity measure of
the probability distributions of the investigating non-commuting
observables,
 $J_x$ and $J_z$,
is bounded by the value $\sqrt{1/2}$. We note that two observables
$A$, $B$ in an $N$-dimensional Hilbert space are said to be
complementary (to each other) if their transformation matrix has the
form \cite{KR}
$$|\langle a_i|b_j \rangle | =N^{-1/2}\qquad (i,j =1,...,N).$$
Complementary observables can be considered as a generalization to
higher dimensions of spin-1/2 orthogonal system \cite{SA}. Hence, we
can proceed similarly also for two complementary observables in a
general N-dimensional Hilbert space.

The concept of Bhattacharyya statistical divergence is quite
independent of quantum mechanics and can be defined in any
probability space. Hence, apart from  the application of the
similarity measure in quantum physics it can also be applied to
theory of the physical measurement. It may serve as a certain degree
of the reliance of a physical measurement. Suppose that in two
different laboratories the probabilities of the decay modes of an
elementary particle are measured. Generally, the different probability distributions of the
 individual decay modes are obtained in each laboratory. To determine the reliance of
the measurement  we can insert the measured probability distributions in formula (1). Here,
the simple rule holds: the large is the value of $S(P_1,P_2)$ the
more reliable is the corresponding measurement.

\begin{acknowledgments}
Partial support by the Grant Agency VEGA No. 2/6087/26 is highly
acknowledged.
\end{acknowledgments}

\end{document}